\documentclass {aa}

\def\simlt{\lower.5ex\hbox{$\; \buildrel < \over \sim \;$}}
\def\simgt{\lower.5ex\hbox{$\; \buildrel > \over \sim \;$}}
\def\today{\ifcase\month\or
January\or February\or
March\or April\or May\or June\or
July\or August\or
September\or October\or November\or
December\fi\space\number\day, \number\year}

\begin{document}
\thesaurus{02(02.13.5; 09.13.2; 12.03.4)}

\title{Quasars and Galaxy Formation}
\author{Joseph Silk\inst{1} and Martin J.\,Rees\inst{2}}
\offprints{J. Silk}
\institute{ Institute of Astronomy, Cambridge, UK,
Institut d'Astrophysique de Paris, France,  and    
 Departments of Astronomy and Physics, University of California,
 Berkeley, CA 94720 USA
\and  Institute of Astronomy, Cambridge, UK}
\date{Received October 9, 1997 / Accepted }

\maketitle

\begin{abstract}

The formation  of massive black holes may precede the epoch that characterises
the
peak of  galaxy formation, as characterized by the star formation
history in luminous galaxies.  Hence protogalactic
star formation may be profoundly affected by  quasar-like
nuclei and their  associated  extensive energetic
outflows. We derive a relation between the mass of the
central supermassive black hole and that of the galaxy spheroidal component,
and
comment on other
implications for galaxy formation scenarios.
\end{abstract}

\section{Introduction}

It is generally assumed that the first objects to form in the universe
 were
 stars.  However this is by no means assured.  There is general agreement among
theorists that, if cosmic structures form hierarchically ('bottom up'), as in
cold dark matter (CDM) models, the  first baryonic clouds  have masses in the
range $10^5-10^6
M_\odot, $ and that the characteristic mass subsequently rises.  But it is
actually not obvious that these clouds would undergo fragmentation into stars.
 Conditions in primordial clouds  differ from those prevailing  in conventional
star-forming clouds that the fate of nearby molecular
 clouds is not a reliable guide. For example, in the absence of
 magnetic
 flux, cloud collapse may have been far more catastrophic than is the case at
the present epoch. A massive disk  could
have rapidly shed its angular momentum  via non-axisymmetric
 gravitational
 instabilities, and become so dense and opaque that it continued to evolve
 as
a single
unit.  At very high redshifts, the inefficiency of atomic and
molecular cooling via H and H$_2$ excitations is compensated by
 Compton
 cooling;
 Compton drag provides an additional mechanism for transferring
 angular
momentum and allowing collapse.

 This outcome seems no less  likely, a priori, than the alternative
evolutionary pathways found in the literature,  according to which
primordial clouds fragment into stars with an initial mass function that varies
between being bottom-heavy, top-heavy or even normal (that is, solar
 neighbourhood-like), depending on the observations that are being interpreted.
The quasar distribution tells us directly that  at least some massive
black
 holes
form early. Indeed,  the quasar comoving density  peaks at $z>2$, and  only
declines at $z>3$ (Shaver {\it et al.} 1996); on the other hand,  the peak of
galaxy formation occurs at $z \approx 1.5$ (Madau {\it et al.}
 1996; Connolly {\it et al.} 1997), although there is uncertainty
about the effects of extinction in leading to an  underestimate
of the galaxy luminosity  at high redshift.

 In fact, the known  quasars, or their dead counterparts,
 are likely to be within the cores 
of at least  30 percent of these galaxies. To see this, note that
combining  the integrated density in
quasar light with  the assumption that quasars radiate at or near the
Eddington limit yields an estimate of typical dead quasar (or black hole)
 mass (Soltan 1982;
Chokshi \& Turner 1992) as  $10^7-10^8\rm \, M_\odot.$
Whether one actually could observe an AGN component in high redshift
galaxies depends sensitively
on the adopted lifetime of the active phase: higher redshift helps.
 The
observed correlation between massive black holes and dynamically hot
galaxies then 
suggests that most hot galaxies, amounting to of order a third of all
galaxies in terms of stellar content, could  contain 
such a  massive black hole (Faber {\it et al.} 1996). One note of
caution would therefore be that dynamically hot galaxies probably form
systematically earlier than most galaxies, and that  if starbursts
characterize their birth, existing high redshift samples
of such objects  may be incomplete.

Nevertheless, while the case remains ambiguous,
we are sufficiently motivated by the
possible implications of a causal connection between quasars and
galaxy formation to
explore in this note  the  consequences of a cosmogonical scenario
in which the first objects to form, at some highly uncertain efficiency,
are supermassive  black holes.  We discuss how, during subsequent mergers, the
holes could grow, and exert a feedback on star formation.

\section{Model}

\subsection{ Early Growth}

We suppose that the initial black holes form 
via a  coherent collapse. This probably implies  $M_{bh} \simgt 10^6\beta
M_\odot,$, with  $\beta \sim 1.$  Formation of lower mass holes would be less
efficient, for at least two reasons.
Primordial clouds of mass less than $\sim 10^9\rm \, M_\odot$
are readily disrupted by supernova-driven winds (Dekel and Silk 1986).
Given the observed efficiency of black hole formation, the  formation of
black holes of mass below $\sim 10^6 \, M_\odot$ is likely to be inhibited.
Moreover
if  the precursor  object forms a supermassive star, there would be
substantial
 mass
 loss.  Since typical first generation
clouds
of primordial CDM  have masses of order $10^6\alpha M_\odot$
where $\alpha\sim 1, $  the total mass going into black holes would be
significant even if they formed in only a small proportion of clouds. Of course
it is also possible that primordial
clouds form smaller black holes which subsequently merge as the
hierarchy develops. However this process involves two additional stages of
inefficiency (via formation and merging), and we regard it
as an improbable pathway.
For a typical $L_\ast$ galaxy to form from  hierarchical merging of
primordial clouds and contain  a supermassive black hole, we require
that the efficiency $f$ at which the
 supermassive
 black hole formed (or equivalently, the inefficiency of fragmentation
in primordial clouds) satisfies $f\simgt 10^{-5}\beta.$

  How small can $f$ be in order for
 the
 consequences
to be of interest? One observes today that many, if not all, galaxies
 contain central supermassive black holes,
and that $M_{bh}=2\times 10^{-3}M_{sph},$ where $M_{bh}$ is the black hole mass
and $M_{sph}$ is the  mass of the spheroidal component
 (Magorrian et al. 1997; Ford et al. 1997; van der Marel 1997). Once
 supermassive black
 holes are formed, the final
 black hole mass is enhanced during
the hierarchical merging process, when dynamical friction and dissipative drag
on gas can drive supermassive black holes into the center of the
developing
 protogalaxy. Mergers provide a continuing supply of gas, and
gas dissipation and accretion feed the central  black hole.
The most detailed numerical simulations of protogalaxy collapse
 with cosmological initial conditions that have hitherto been
performed
 demonstrate that angular momentum transfer is highly
effective (cf. Navarro and Steinmetz 1997). Baryons
 collapse to form
 a dense, massive central clump at the resolution limit of the simulations,
 rather than a centrifugally-supported disc on a galactic scale. This line of
thought at least supplies a motivation for exploring (and constraining) the
hypothesis that black
 hole formation and growth, rather than star formation, characterizes
the
 earliest stages of galaxy formation. We will argue that the
value of $f$ self-regulates so as to approximately satisfy the
 observed correlation.

\subsection{ Quasar Winds}

Massive black holes, whenever fuelled at a sufficient rate, would display
quasar-like activity. For brevity we term such objects 'quasars' -- noting,
however, that the events we are discussing may occur at higher redshifts than
the typical observed quasars. An explosion model whereby outflows from early
quasar-like objects led to  cooled shells which fragmented into galaxies was
originally developed by Ikeuchi (1981; {\it cf.}
 also Ostriker  and Cowie 1981); this idea has,
however,  fallen from favour as
the principal mode for galaxy formation because post-shock Compton
 cooling would
lead to excessive  spectral distortions of the cosmic microwave background
spectrum.  We consider here the (more localised) effects  on the  gas  within
the protogalaxy in which the quasar is embedded.

 The effect of a protogalactic
wind may be estimated as follows.
We model a protogalaxy  as an isothermal sphere
 of cold dark matter that
contains gas fraction $ f_{gas} $ with density $ \rho =
\sigma ^2 / 2 \pi G r^2 $, constant velocity
dispersion $ \sigma $, and mass $ M (<r) = 2 r \sigma
^2 /G. $    In massive halos,
$ T
\approx {\sigma ^2  m_p / 3k } = 4 \times 10^6{\rm  K}
 \left( {  \sigma / 300{\rm km  \, s^{-1} }  } \right)^2  .$
 A sufficiently intense wind
from the central quasar can sweep up the gas into a shell,
 and push it outwards
at constant velocity
$$v_s =\left(f_wL_{Edd}8\pi^2G\over{f_{gas}\sigma^2}\right)^{1/3}.$$
In this expression, the mechanical ({\it i.e.}
 wind)  luminosity is taken to be a
fraction $f_w$ of the Eddington luminosity $L=4\pi
GcM_{bh}\kappa^{-1}=1.3\times10^{46}M_8\rm \, ergs\, s^{-1},$
Note that
$f_w\equiv \dot M_{out}v_w^2/L_E=\epsilon^{-1}(v_w/c)^2\approx 0.01,$
where $\epsilon=L_E/\dot M_{inf}c^2 $ is the radiation efficiency.

Expulsion of this shell requires that its velocity  should exceed
 the escape velocity from the protogalaxy: {\it i.e.} $v_s>\sigma.$
The condition for this to be the case is that
$$M_{bh}>\alpha{\sigma^5\kappa\over G^2c} =8\times 10^8\gamma(\sigma/500\,
\rm km\,s^{-1})^5\, M_\odot,\eqno(1)$$
where $\sigma_{500}\equiv {\sigma/500\,
\rm km\,s^{-1}} $ and $\gamma^{-1}\equiv
32\pi^3f_w/ f_{gas}\sim 1.$ If for example  $\gamma\sim 1$, black
holes
 could in
principle  eject all the material from their host galaxies when their masses
exceed $\sim 10^7 M_\odot.$
If the situation were indeed spherical, then, even if the central source
switched off,  the outflow would  continue to expand into the intergalactic
medium for up to a Hubble time before stalling  due to the
 ambient pressure. The
shell velocity decreases via
an  explosive outflow  according to
$$v_s=330  (E_{62}/\Omega_{0.05})^{1/5}(1+z)^{3/10}\rm \, km\, s^{-1},$$
for an explosion energy, equal to the kinetic energy at breakout,
of $10^{62}E_{62}$ ergs and an intergalactic medium density equal to
5 percent of the Einstein de Sitter density (with $H_0=60\rm \, km\, s^{-1}\,
Mpc^{-1}$), as compared to the
binding energy of the gas in a massive protogalaxy
of $\sim 10^{60}-10^{61}$ erg.
The ambient pressure is not high enough to halt the shell, initially
 moving at breakout at a
 velocity of $\sim 500\,\rm km\,s^{-1}$, until $z\simlt 1.$
The shell radius
 is $R_s=4.6(E_{62}/\Omega_{0.05})^{1/5}(1+z)^{-6/5}\rm Mpc, $
 and the fragment mass is  (Ostriker  and Cowie 1981)
$$M_f=7\times 10^{10}
a_{10}^4E_{62}^{-1/5}\Omega_{0.1}^{-4/5}(1+z)^{-9/5}
M_\odot,
$$
where $10 a_{10}{\rm km\,s^{-1}}$ is the sound velocity within the shell.

    How effective this expulsion would actually be,   depends on the
 geometry of
the outflow and on
the degree of inhomogeneity of the protogalactic gas.  Realistically, the
outflow may be directional (probably bipolar)
 rather than spherically-symmetric.
Some gas may survive
and even be overpressured by double radio lobes
to form stars (Begelman and Cioffi 1989)

 It is even possible (Natarajan et al 1997) that swept-up material,
 after cooling
down  into a dense shell, may fragment into a class of dwarf galaxies; these
would differ from normal dwarfs in not being embedded in  dark halos, and
consequently be more  liable to disruption by massive star formation.
  Hence  a high-mass black hole has two contrasting effects. It inhibits star
formation in its host halo by blowing gas out; on the other hand, the ejected
gas may eventually pile up in a cool shell that breaks up into small galaxies.

\subsection{Protogalaxy Core }

 The implications of  (1)  are that a massive hole, if it continues to emit at
close to the Eddington rate,  could  expel gas completely from its host galaxy.
(The amount of gas that has to be accreted is in itself negligible in this
context: each unit mass of  gas accreted into the hole  can release enough
energy to expel of order $(c/v)^2$ times its own mass from the far  shallower
potential well of the halo.)  However, the fuelling demands a continuing
accumulation  of gas in the centre (probably supplied by hierarchical merging).
We can therefore  interpret (1) as setting an upper limit to the mass of a hole
that can exist in a galaxy where star formation can proceed efficiently.

   Expression (1) gives a relation between a black hole and its surrounding
halo.   If a significant fraction of the Eddington luminosity   emerged in
'mechanical ' form, this gives the criterion that the energy liberated on the
dynamical timescale of the halo is equal to the gravitational binding energy.
The same expression can be derived by a different argument. The maximum rate at
which gas can be fed towards the centre of a galaxy (as the outcome of mergers,
etc) is   $\sim\sigma^3/G,$
where $\sigma$ is the velocity dispersion  in the merged protogalactic
system.  A quasar could   expel all this gas from the galactic
 potential well on
a dynamical timescale if its mass exceeded a critical value obtained by
requiring that
$ \sigma^5/G=4\pi GcM_{bh}^{cr}/\kappa.$

This is
indistinguishable from the observed relation
between the masses of central holes
and those of their host spheroids ({\it e.g.} Magorrian {\it et al.}
1997).
   The observed
relation  has considerable scatter but hints at a dependence of
 black hole on  spheroid mass that rises more rapidly than linearly:
{\it e.g.}
 the Milky Way has a central black hole mass of $2.5\times 10^6M_\odot,$
whereas
 M87, with a spheroid mass that
is only $\sim 100$ larger than that of the Milky Way,
 has a central black hole of mass $4\times 10^9M_\odot$.

We therefore hypothesize
 that, in the merger process leading to the formation of
typical galaxies,  the hole mass  stabilises near the critical value.
Hierarchical merging, augmented by continuing black hole growth, helps maintain
the black hole-to-spheroid mass ratio at the self-regulation level.
  We suggested in section 1, however,  that in the first bound systems gas may
accumulate into a single compact unit and evolve into a supermassive
 hole.
 If
single black holes are indeed favoured over star formation in this way, the
holes in these first systems would be far above the 'critical' value appropriate
to such small halos with  low velocity dispersions: in other words $M_{bh}\gg
M_{bh}^{cr}.$
 Star formation would then be inhibited (except in a disc close to the central
quasar)  until, via mergers, the halos had become large enough to bring the mass
of the actual central hole below the critical mass (which, as we have seen,
grows faster than linearly with halo mass). Thereafter, the scaling would be
maintained.

If.  after mergers had formed high-mass halos, the central hole were still
above the critical mass,   the implications  would be  ominous for galaxy
formation. Such a system would end up as a supermassive black hole
embedded in a
low surface brightness galaxy. Around the central host quasar, one would expect
to find a
cavity of hot gas. Compton cooling at high redshift will tend to
quench any associated extended radio emission. If large radio sources were
responsible for the hot gas bubbles, then the geometry is
not a sensitive issue. Indeed, double lobes are as effective as
spherical
outflows
and
are more reminiscent of the geometry
of the associated hot gas
bubbles with the similar  energetics of reported Sunyaev-Zeldovich decrements
that have  no apparent associated galaxy cluster but with possibly associated
quasars (Jones {\it et al.} 1997; Partridge {\it et al.}
1997). Natarajan {\it et al.} 1997) argue that
one might expect to detect a shell
of newly formed Magellanic irregular-type galaxies at the periphery of
such
 bubbles.

\section{
IMPLICATIONS FOR GALAXY FORMATION AND SUPPRESSION}

Near the hole, the ionizing radiation certainly suppresses star formation
in normal molecular clouds, which, if of density $n$,
 survive only at a distance greater than\\
 $\sim 10 (L_{46}/n)^{1/2} \rm \, kpc$ from the central quasar.
The ultraviolet flux is  effective at destroying   $H_2$  molecules
 produced via $H^-$
formation to much  larger distances,
{\it e.g.} the $H^-$   photodissociation rate is  $\sim
10^{-10}L_{46}r_1^{-2}
\ \rm s^{-1}$ at $r_1\, \rm  Mpc$ from the quasar whereas the rate of the compet
ing
 $H_2$ formation process is $\sim 10^{-9}n\ \rm s^{-1}$ (Tegmark {\it et
al.} 1997).
This would
tend to suppress  formation of dwarf galaxies out to
 $\sim 0.3 n^{-1/2}\ \rm Mpc$,
 If the radiation extends to the x-ray band, there is  however  a narrow regime
where the hard ionizing flux, by maintaining a fraction of free electrons deep
inside the cloud,  may
actually enhance star formation by stimulating $H^-$  formation (Haiman, Rees
and Loeb 1996).

There are several further  noteworthy consequences of the hypothesis that
supermassive black holes form within the first  subgalactic structures that
virialise
at high redshift,
and are in place  before
most galactic stars have formed. AGN activity
 generates outflows that can interact dynamically  with the
surrounding protogalactic gas as well as provide a possible early flux
 of hard ionizing photons. Star formation in the   accretion
disk surrounding the broad emission line region of the AGN is likely
to proceed under conditions very different from those
 encountered even in starbursts. Star-forming
clouds at distance $r_{pc}\, \rm pc$ from the supermassive black hole
 must be dense enough to avoid tidal disruption,
or $n\simgt 10^9M_{8}r_{pc}^{-3}\rm cm^{-3}.$
One infers that the Jeans mass ($\propto a_s^3/\rho^{1/2}$)
is reduced to  stellar scale, and also that
 the specific angular momentum of such a clump is reduced
 (as $\propto a_s^2/\rho^{1/2}$), relative to values in conventional
star-forming clouds.

Suppose that the circum-quasar accretion disk is dense enough to be
self-shielding and to be predominantly molecular. If 
enriched to near-solar abundance level, CO molecules and other species 
provide important cooling and should result in a temperature of order
$100\,\rm K.$ One then infers 
that two of the classical barriers
 to star formation are likely to
 be
 overcome. Hence any gas which falls close enough to the centre to be
part of an accretion disc should convert efficiently into stars.   Outflow
 from these stars that form under the gravitational influence of the hole
provides a source of metal enrichment, and may thereby 
 stimulate more widespread  star formation in a protogalaxy. Magnetic flux will
be ejected, the accretion disk  providing the conditions for
an efficient dynamo. The subsequent turbulent mixing by supernova
remnants
 and stretching by differential rotation could provide a possible
 origin for the interstellar magnetic field.

There are also interesting consequences for the intergalactic medium (IGM),
since quasar-driven winds could readily eject enriched material from the shallow
potential wells that characterize the
earlier
 stages of hierarchical clustering.  This could account for the apparent
presence of up to  1 percent of the solar abundance of heavy elements even at
high z. 
If this is the case, a time delay is likely between
 the epoch of most quasar activity and the epoch of  the bulk of star formation
in the universe. One observes a time delay corresponding to
 $\delta z\approx 1$ : studies of quasars
have confirmed a peak in the quasar number density  between $z=2$ and
 3, and
 the
 star formation history of the universe is found to peak between
$z=1$ and 2.

The cores of  present-day galaxies may carry traces of the black hole
 merging
history.
Megers and accretion will allow the black hole to grow to
 the critical value as
determined by equation (1).
Our key prediction is the relation between central
black hole mass and spheroid mass.

Luminous elliptical cores are best explained by heating associated with
binary black hole decay following a major merger
 (Magorrian {\it et al.} 1997). The present model envisages that this
 process operates at all stages of the hierarchy
 but only towards the end would the gas be
 mostly exhausted. Perhaps this would help explain the transition
with increasing luminosity  from coreless spheroids
to hot galaxies with cores.

\bigskip

We  acknowlege helpful discussions with M. Haehnelt and P. Natarajan.  
The research of JS has been supported in part by grants from
NASA and NSF, and that of MJR by the Royal Society. JS  acknowledges with
gratitude the hospitality
of the Institut d'Astrophysique de Paris  where he is a Blaise-Pascal
 Visiting Professor, and the Institute of Astronomy
at Cambridge, where he is a Sackler Visiting Astronomer.

\bigskip

\end{document}